\documentclass{article}

\usepackage[nonatbib,preprint]{neurips_2023}
\usepackage[numbers]{natbib}
\usepackage{placeins}
\usepackage{lscape}
\usepackage{longtable}
\usepackage{subcaption}

\usepackage[utf8]{inputenc} 
\usepackage[T1]{fontenc}    
\usepackage{amsmath}
\usepackage{amssymb}
\usepackage{graphicx}
\usepackage{hyperref}       
\usepackage{url}            
\usepackage{cleveref}
\usepackage{booktabs}       
\usepackage{amsfonts}       
\usepackage{nicefrac}       
\usepackage{microtype}      
\usepackage{xcolor}         
\usepackage{enumitem}
\usepackage{float}
\usepackage{MnSymbol}       

\newcommand{\admet}{ADME(T)}

\title{Current Methods for Drug Property Prediction \\ in the Real World}

\author{%
  Jacob Green \\
  DeepMirror \\
  \texttt{jacob@deepmirror.ai} \\
  \And
  Cecilia Cabrera Diaz \\
  DeepMirror \\
  \texttt{cecilia@deepmirror.ai} \\
  \And
  Maximilian A.~H.~Jakobs \\
  DeepMirror \\
  \texttt{max@deepmirror.ai} \\
  \And
  Andrea Dimitracopoulos \\
  DeepMirror \\
  \texttt{andrea@deepmirror.ai} \\
  \And
  Mark van der Wilk\thanks{Work done while at DeepMirror.} \\
  Imperial College London \\
  \texttt{m.vdwilk@imperial.ac.uk}
  \And
  Ryan D.~Greenhalgh\thanks{Corresponding author.} \\
  DeepMirror \\
  \texttt{ryan@deepmirror.ai} \\
}

\begin{document}

\maketitle

\begin{abstract}
Predicting drug properties is key in drug discovery to enable de-risking of assets before expensive clinical trials, and to find highly active compounds faster. Interest from the Machine Learning community has led to the release of a variety of benchmark datasets and proposed methods. However, it remains unclear for practitioners which method or approach is most suitable, as different papers benchmark on different datasets and methods, leading to varying conclusions that are not easily compared. Our large-scale empirical study links together numerous earlier works on different datasets and methods; thus offering a comprehensive overview of the existing property classes, datasets, and their interactions with different methods. We emphasise the importance of uncertainty quantification and the time and therefore cost of applying these methods in the drug development decision-making cycle. We discover that the best method depends on the dataset, and that engineered features with classical ML methods often outperform deep learning. Specifically, QSAR datasets are typically best analysed with classical methods such as Gaussian Processes while ADMET datasets are sometimes better described by Trees or Deep Learning methods such as Graph Neural Networks or language models. Our work highlights that practitioners do not yet have a straightforward, black-box procedure to rely on, and sets the precedent for creating practitioner-relevant benchmarks. Deep learning approaches must be proven on these benchmarks to become the practical method of choice in drug property prediction.
\end{abstract}

\section{Introduction}
\label{sec:goals}

In the high-stakes field of drug discovery, where the development of a new pharmaceutical often takes 10 to 15 years and requires investment exceeding \$1bn before it is available to the public, the need for more efficient strategies is apparent. Predictive models are emerging as a compelling solution for the virtual screening of compounds, which can reduce the need for costly synthesis or execution of expensive assays during the hit-to-lead or lead optimisation phases of drug discovery, by helping focus experiments on compounds likely to be effective, based on previously gathered data.

To discover such new compounds, one has to both exploit known effective structures and explore ones further afield from previously measured ones \cite{hennig2022probabilistic}. Therefore, practitioners need to both predict properties and quantify prediction uncertainty to identify structures that have good properties with low uncertainty (exploit) and those that have high uncertainty but potentially even better properties (explore). Robust uncertainty quantification provides a more nuanced understanding of prediction outcomes, allowing practitioners to integrate these predictions seamlessly into other decision-making frameworks. Therefore, the development and application of effective uncertainty quantification methods can significantly enhance the efficiency and cost-effectiveness of the drug discovery process.

Traditional methodologies such as Random Forests (RF) and Support Vector Machines (SVM) have demonstrated efficacy in molecular property prediction \cite{molace, admet_trees1, admet_trees2, admet_trees3, xgboost_tdc}, but the relative performance of a specific method compared to the others varies across datasets. The development of deep learning, specifically Graph Neural Networks (GNNs), has further complicated choosing the best method \cite{wu2018moleculenet, veličković2018deep, alphaleegnn}. While GNNs show potential, their performance gain is inconclusive when compared across various studies and compared to classical models \cite{molace, couldgnnsbebetter}. Pre-training techniques using GNNs \cite{google_pretrain_gnn, infomax_maybe_pretrain_helps}, language models, \cite{mtlbert, chithran2020chemberta}, and 3D structures \cite{stärk20223d} have shown promise, but have not been sufficiently benchmarked against competing methods to draw conclusions about general performance. So, in spite of the development of many methods, there has never been a comprehensive comparison, and it is therefore difficult to make a fully informed decision when selecting the most appropriate machine learning model for molecular property prediction. Additionally, these methods rarely quantify uncertainty leading to a significant gap in the literature. Our work aims to address this gap and navigate these inconsistencies, thereby providing a more comprehensive understanding of selecting the optimal method for molecular property prediction with  uncertainty quantification. Hence, we endeavour to answer the following research questions:

\begin{enumerate}[label=\textbf{RQ\arabic*.}]
    \item From a large collection of classical methods (GPs, SVMs, RFs etc.) and modern methods (GNNs and language models), does one consistently outperform across a variety of datasets?
    \item Can groups of datasets be identified where a specific method routinely outperforms all others?
    \item How can one quantify and validate the uncertainty of different methods?
    \item What are the trade-offs between computational time and predictive performance in the application of these methods?
\end{enumerate}

Answering these questions provides actionable advice to practitioners aiming to build predictive models for molecular datasets. In addition, it provides a clear assessment of how modern deep learning techniques compare to classical methods for these problems, and sets a standard that needs to be reached for a method to become a clear method of choice.

\section{Problem Setting: Drug Property Prediction and Datasets}
\label{Datasets}
During drug development, two classes of properties are generally predicted: Absorption, Distribution, Metabolism, Excretion, and Toxicity (\admet{}) and Quantitative Structure-Activity Relationships (QSAR). \admet{} properties are often optimised to fall within a given range in order for them to be safe and effective for patients. QSAR refer to properties that are drug specific, such as the inhibitory effect on a target enzyme, or the binding affinity to a particular receptor that is linked to a disease. Being able to accurately predict these properties prior to carrying out laboratory testing is crucial to reduce the time and resources needed to bring safe and effective new drugs to patients. There are several public repositories with datasets of \admet\ and QSAR properties that can be used for developing and benchmarking predictive models, the most prominent of which are the Therapeutics Data Commons (TDC) \cite{Huang2021tdc, Huang2022artificial}, MoleculeNet~\cite{wu2018moleculenet}, and OPERA \cite{mansouri2018opera}.

Overall, the existing literature presents an incomplete, or even conflicting picture of the performance of GNNs and classical ML methods in drug property regression problems, as most earlier works test against a single benchmark and only with subsets of machine learning methods (\cref{past_lit}).  While this can be expected to an extent given practical constraints on running evaluations, it complicates selecting a model for practitioners in science and business, who need an effective solution for their application. We believe that providing results on MoleculeNet is not sufficient to provide an idea of performance across drug property prediction classes. Many other papers only compare subsets of classical methods, leaving out some candidates (such as Gaussian Processes) altogether. Uncertainty in predictions is a vital element in de-risking decision making during drug discovery, yet only a small subset of research studies have been conducted to evaluate uncertainty \cite{ensembles, alphaleegnn, griffiths2022gauche}. 

We aim to address these gaps by comparing a wide set of methods, over a wide range of datasets, while also evaluating uncertainty quantification, to gain the first complete picture of how such problems can be tackled at scale. We select as wide a range of datasets as possible, but restrict ourselves to regression datasets, as they are more meaningful and realistic in real-world settings because:
\begin{enumerate}
    \item Most molecular property prediction classification tasks are derived by thresholding regression problems, resulting in loss of information.
    \item When converting a problem into a classification task, practitioners must choose a subjective threshold which may differ among individuals, even within the same organisation. This also influences performance metrics and model considerations (e.g.~through class imbalance).
\end{enumerate}
In addition, we take care to include difficult datasets with ``activity cliffs'', where minor alterations in molecular structure can cause significant property changes (by several orders of magnitude). We obtain these from the curated ChEMBL datasets from MoleculeACE \cite{molace}.

For all benchmark datasets \cite{molace, wu2018moleculenet,Huang2021tdc, mansouri2018opera}, we adhered to the default split as reported in the original benchmark. For the remaining datasets, we conducted train/test splits using a MaxMin splitter \cite{deepchem} to ensure that compounds in the test set were structurally dissimilar from those in the training set. In total, we benchmarked 184 machine learning methods on 44 regression datasets, covering two parent property classes: QSAR (32) and \admet{} (12). See \autoref{tab:datasets_summary} for a detailed overview.

\section{Current Methods and Challenges in Drug Property Prediction}
\label{litreview}
\subsection{Method Overview}
\label{sec:method-overview}
\paragraph{Input generation}
A common representation of molecules is the Simplified Molecular Input Line Entry System (SMILES), a compact notation that uses ASCII strings to represent chemical structures in a readable format. All benchmark datasets describe individual molecules using SMILES. In order to construct predictive models for molecular structures, these structures must first be translated into a numerical form, with three of the most popular approaches being:
\begin{enumerate}
    \item \textbf{Cheminformatics molecular descriptors} that are designed using human understanding. These are constructed based on an extensive range of precalculated properties, encompassing geometrical, electronic, and thermodynamic attributes. Some techniques stop here, while others examine each atom in the molecule and its local environment up to a defined radius or bond distance, then formulate a unique identifier (fingerprint bit) for each atom and its corresponding environment, typically via a hashing procedure.
    \item \textbf{Graph representations} where atoms are depicted as nodes and bonds as edges. The encoding of atoms and bonds within nodes and edges can be achieved in various ways, such as one-hot encodings or vectors characterised by physical properties. With graph representations, it is possible to apply graph-based machine learning algorithms to predict properties of the molecule directly from its structure.
    \item \textbf{Tokenisation} of SMILES strings, by segmenting the string into smaller chunks or tokens, which typically involves breaking down the string into individual atoms, bonds, and ring closures. Tokenised SMILES can be inputs to language models, e.g. Transformers.  
\end{enumerate}

\vspace{-0.3cm}
\paragraph{Classical vs. Deep Learning} Different machine learning models can work with different input formats. In this paper, we distinguish ``classical'' machine learning methods, and deep learning methods. Classical methods require inputs to be real-valued vectors, and do little in the form of automatic feature design or extraction. As such, they are highly reliant on descriptive features being designed \textit{before} being passed to the machine learning method. Deep learning, on the other hand, operates on representations of the data that are less pre-processed, like the graph representation or tokens, which allows feature representations to be learned in a data-driven manner, thus overcoming the limitations of fixed features. Specialised architectures have been developed to deal with graph or sequence structured inputs (GNNs \cite{alphaleegnn} and Transformers \cite{chithran2020chemberta, mtlbert, ahmad2022chemberta2}). Deep learning can also be applied to extract features, which can then be used as inputs to classical methods, as explained below.

\vspace{-0.3cm}
\paragraph{Classical methods} The fixed features that classical methods rely on can be human-designed, or learned in a self-supervised or unsupervised manner using deep learning. Features learned from data always generate the same output for a given input (deterministic). There are a variety of human-designed features in the cheminformatics literature, many of which are close variations of the two that we consider: \textbf{ECFP} \cite{rogers2010extended} and \textbf{Mordred} \cite{moriwaki2018mordred}. We do not consider the ``PubChem'' fingerprint \cite{pubchem_fingerprint}, as it was shown to have less feature importance for tree models compared to other cheminformatics descriptors \cite{xgboost_tdc}. Learned feature descriptors, such as \textbf{Mol2vec} \cite{jaeger2018mol2vec}, were first inpsired by Natural Language Processing (NLP), with newer ones using transformers, like \textbf{ChemBERTa-2} \cite{chithran2020chemberta, ahmad2022chemberta2} and \textbf{MTL-BERT} \cite{mtlbert}. Finally, self-supervised learning on graphs \cite{veličković2018deep} has been used to create \textbf{InfoMax2D} \cite{infomax_maybe_pretrain_helps} and \textbf{InfoGraph3D} \cite{stärk20223d} features, which we also test. Each of these feature extractors can be used together with a variety of classical machine learning methods, such as \textbf{SVMs} \cite{cortes1995support}, decision trees (\textbf{Random Forests (RF)} and \textbf{Light Gradient Boosting Machine (LGBM)} \cite{lgbm}), \textbf{K-Nearest Neighbours (KNN)}, \textbf{Gaussian Processes (GP)} \cite{gpml}, and \textbf{Multilayer Perceptrons (MLP)}.

\vspace{-0.3cm}
\paragraph{Deep learning methods} Deep learning methods such as GNNs have surged in popularity in drug discovery due to their ability to directly work on the 2D graph representation of molecular structure. In theory, this should allow the most effective features to be learned from data. Here, we included a number of architectures: \textbf{Directed Message Passing Neural Networks (D-MPNN)} \cite{yang2019analyzing}, \textbf{Graph Convolutional Networks (GCN)} \cite{kipf2016semi}, \textbf{Attentive Fingerprint (A-FP)} \cite{xiong2019pushing}, and \textbf{Graph Attention Networks (GAT)} \cite{velivckovic2017graph}. Other deep learning methods such as LSTMs and CNNs have previously been assessed in this domain \cite{molace}; we chose not to include them in our benchmark as these model architectures have been superseded by transformer-based models in recent years.

\subsection{Conclusions and Recommendations from Past Literature}
\label{past_lit}
Multiple publications have compared machine learning methods for molecular property predictions. Here, we outline the most important works and their findings.

Graph Neural Networks are currently widely studied in the hope that feature learning directly on the molecular graph leads to better performance. Recent works \cite{alphaleegnn, wu2018moleculenet, yang2019analyzing} show continuous improvement in GNN performance on MoleculeNet regression datasets \cite{wu2018moleculenet}. GNNs are often only compared to other GNNs \cite{stärk20223d}, or on a single benchmark dataset \cite{li2017gnnsrawesome_1}, preventing general conclusions in a wider context. \citet{yang2019analyzing} compare GNNs to classical methods (RFs and MLPs -- a subset of those in \cref{sec:method-overview}) on MoleculeNet, and find that in 1 out of 3 regression datasets GNNs perform best. \citet{couldgnnsbebetter} on the other hand, report the same GNN method being outperformed by SVMs. In \cite{wu2018moleculenet}, where MoleculeNet was originally introduced, it was found that XGBoost performed the best. It is difficult to draw conclusions across papers where different subsets of models are compared each time. More recently, pre-trained deep learning models \cite{jaeger2018mol2vec,mtlbert,chithran2020chemberta, ahmad2022chemberta2} showed promise, with e.g. \citet{ahmad2022chemberta2} demonstrating improvements in performance over classical models in 3 of 4 datasets.

Papers centred on a particular application, such as TDC, often perform more comprehensive benchmarks using classical methods. \citet{Huang2021tdc} introduce the TDC datasets and benchmark molecular descriptors and graph based models, with different methods working best in different situations. Later, \citet{xgboost_tdc} showed XGBoost models were highly performant across a range of the TDC benchmarks. Similarly, \citet{molace} introduced the MoleculeACE datasets, and showed classical models (SVMs) combined with ECFP outperforming all deep learning models. \citet{griffiths2022gauche} suggest that Gaussian processes can also be performant, but only compare to simple (Bayesian) neural networks.

Finally, only a small subset of studies evaluate methods that explicitly quantify uncertainty \cite{ensembles, alphaleegnn, griffiths2022gauche}. For practitioners, uncertainty around model predictions is a key metric when prioritising molecules for synthesis and experimentation. There are numerous ways to encapsulate uncertainty in models. For instance, variability among ensemble members \cite{ensembles}, leveraging inherent variance in GPs, \cite{griffiths2022gauche}, and Dropout Variational Inference or Stein Variational Gradient Descent for GNNs \cite{alphaleegnn}. However, the metrics used for uncertainty quantification tend to diverge significantly between studies so that no clear picture has emerged, yet.

\section{Experimental Procedures}
\label{methods}
\subsection{Set-up and configuration}
In our study, we developed a Python-based pipeline for comprehensively comparing various methods and models for drug property regression tasks, addressing the inconsistencies and gaps in the current literature (see \cref{sec:method-overview}). We aimed to use each model as it would be used in a real setting, giving each model the highest chances of scoring top on a given dataset (those discussed in \cref{Datasets}). Therefore, each model is provided with training data alone, and the handling of the data is optimised for the model in question. For instance, GPs in real-world applications do not require validation data and can therefore use all the training data to build a model, while a GNN training loop requires validation data to prevent overfitting, thereby sacrificing some training data. In the context of real-world applications, we consider this procedure to be the fairest way to compare models. 

As discussed in \cref{past_lit}, only a small number of papers investigate uncertainty quantification. The simplest way to extract an uncertainty measurement for a given prediction is to train an ensemble of models and measure the variance in the models' predictions. Therefore, in addition to training single models, we also trained a corresponding ensemble of models for each model type. This enabled us to estimate uncertainty while potentially improving the associated point prediction. Despite ensembles improving uncertainty estimation, their predictive variance may be \emph{uncalibrated}, and not match the variance of the future testing data \cite{palmer2022calibration}. 
We also consider further improving the uncertainty estimates of ensembles through a post-hoc calibration procedure, at the cost of requiring an additional holdout calibration dataset. Similar to models that require validation data at training time, calibrated models sacrifice training data. This reduces the amount of training data available for optimising model parameters, possibly presenting a trade-off between inhibited model performance and improved uncertainty estimates.

This resulted in any base model (SVM, GP, GNN, etc.) having 4 distinct model flavours: Single Model, Ensemble Model, Calibrated Single Model and Calibrated Ensemble Model. Our model library included SVM, RF, KNN (sklearn, 1.0.2), LGBM (\cite{lgbm} 3.3.5), GP (GPyTorch \cite{gardner2018gpytorch}, 1.8.1), MLP, A-FP, GAT, GCN, and D-MPNN (DeepChem \cite{deepchem}, 2.7.1). Each model was paired with featurisers such as graphs \cite{deepchem} (MGC and D-MPNN for GNNs only), Extended-Connectivity Fingerprints (ECFP, radius 2, number of bits 1024), Mordred \cite{moriwaki2018mordred}, and Mol2Vec \cite{jaeger2018mol2vec}. We also used pre-trained graph models and transformers as featurisers. InfoMax2D was trained in-house based on previous literature \cite{veličković2018deep, infomax_maybe_pretrain_helps}. InfoGraph3D was pre-trained on the Drugs database as outlined in \cite{stärk20223d}. ChemBERTa-2 was used directly from Hugging Face \cite{chithran2020chemberta} (ChemBERTa-77M-MTR). MTL-BERT was trained on 2 million compounds from ChEMBL \cite{gaulton2017chembl} as in \cite{zhang2023mtlbert}. 

Ensemble models were trained with 3 base models. We conducted a random search with 3-fold cross-validation over a hyperparameter search space of 20 hyperparameter combinations to optimise model performance. Deep learning models were trained for 300 epochs with an early stopping schedule of no new lowest validation loss for 50 consecutive epochs. Models requiring validation and/or calibration sets used an 80/20 random split. For each dataset, models were trained using all compatible model-featuriser sets. In total 184 descriptor and model combinations were trained for each dataset. A cluster of 40 NVIDIA T4 GPUs was used for parallelised training across datasets, with models for each dataset being trained on a single GPU.

\subsection{Uncertainty Quantification, Evaluation, and Calibration}
\label{subsec:how2calibrate}

We are interested in quantifying uncertainty because this helps decision makers balance risk. For example, when searching for a compound with desired properties, experiments in unexplored regions (with high uncertainty) should be balanced with experiments in known promising regions. This procedure, even when performed with human input, is somewhat like Bayesian optimisation \cite{hennig2022probabilistic}, and uncertainty is crucial for balancing exploration and exploitation. If a method's predictive distribution matches the frequency of outcomes in the future, optimal decisions can be made.

For evaluation, we use \emph{proper scoring rules} \cite{gneiting2007strictly}, which have the property that they are minimised only when the predictive distribution matches the test set distribution. This allows the quality of a probabilistic forecast to be summarised in a single number that combines accuracy and the quality of uncertainty estimate. This is unlike pure calibration metrics (e.g.~expected calibration error, or ECE, for classification), which can be maximised even if predictions have the accuracy of a random guess.
We choose the ``negative log predictive density'' (NLPD) scoring rule, which has a clear interpretation as the return from a repeated betting game where resources need to be spread across different bets \cite{kelly1956new}. We chose this as a proxy for spreading resources among developing different compounds.

Our predictive distributions quantify both epistemic (model uncertainty) and aleatoric (inherent system or process uncertainty) uncertainty \cite{der2009aleatory}. An estimate of aleatoric uncertainty can be found for all methods by calculating the mean-squared error on the training data, and using this as the variance in a noise model, i.e.~given the prediction $f(x^*)$ we model our observation $y^*$ as
\begin{equation}
    p(y^* | f(x^*)) = \mathcal{N}\left(y; f(x^*), \sigma^2\right)\,, \qquad\qquad \sigma^2 = \frac{1}{N}\sum_{n=1}^N \left(f(x^{\text{train}}_n) - y^{\text{train}}_n\right)^2 \,.
\end{equation}
For ensemble models, we estimate the epistemic uncertainty in the prediction again as a Gaussian distribution, with mean and variance given by that across the ensemble components:
\begin{gather}
    p(f(x^*)|\text{data}) \approx \mathcal{N}\left(f(x^*); \mu(x^*), \nu(x^*)\right)\,, \\
    \mu(x^*) = \frac{1}{K}\sum_{k=1}^K f^{\text{ens}}_k(x^*) \,, \qquad\qquad \nu(x^*) = \frac{1}{K}\sum_{k=1}^K \left(f^{\text{ens}}_k(x^*) - \mu(x^*)\right)^2 \,.
\end{gather}
Gaussian processes provide a direct calculation of $p(f(x^*)|\text{data})$ \cite{gpml}. Using this predictive distribution, we can calculate the NLPD score as
\begin{align}
    \text{NLPD} = - \frac{1}{N_{\text{test}}}\sum_{n=1}^{N_{\text{test}}} \log \mathcal{N}\left(y^{\text{test}}_n; \mu(x^{\text{test}}_n), \nu(x^{\text{test}}_n) + \sigma^2\right) \,.
\end{align}

In cases where predictions are systematically over/underconfident, also consider calibrating the models on a hold-out calibration set. To do this, we scale the predicted mean and variance to make the scaled residuals $(y_n^{\text{test}} - \mu(x^{\text{cal}}_n)) / \sqrt{\nu(x_n^{\text{cal}}) +\sigma^2}$ have zero empirical mean and unit empirical variance. When considering only rescaling, this minimises NLPD on the calibration set.

\section{Experimental Results and Discussion}

\subsection{Which machine learning model, across the broad spectrum of existing models, consistently exhibits superior performance in a variety of datasets?}
\label{q1_results}
We trained all of our uncalibrated descriptor-model combinations on each of our QSAR and \admet{} datasets (\cref{litreview}). We compared model performance using the root mean squared error normalised by the interquartile range of the training data (NRMSE $ = \frac{RMSE_{test}}{IQR_{train}}$). The NRMSE enables direct comparison of performance across datasets and can be interpreted as the "predictive resolution" of a model, i.e., the resolution to which accurate predictions can be made. The IQR was chosen as the normalisation constant due to its robustness to outliers compared to the standard deviation. The training data can be viewed as a representative sample from the true underlying population in both feature and target space. For (NRMSE $<1$) the mean error is less than the spread of the prediction target. Conversely, an (NRMSE $>1$) indicates that the expected error in the prediction is greater than the IQR of the training data indicating poor performance. A numerical summary of experimental results is given in \autoref{tab:experimental_results}.

\begin{figure}[h]
\centering
\includegraphics[width=1\textwidth]{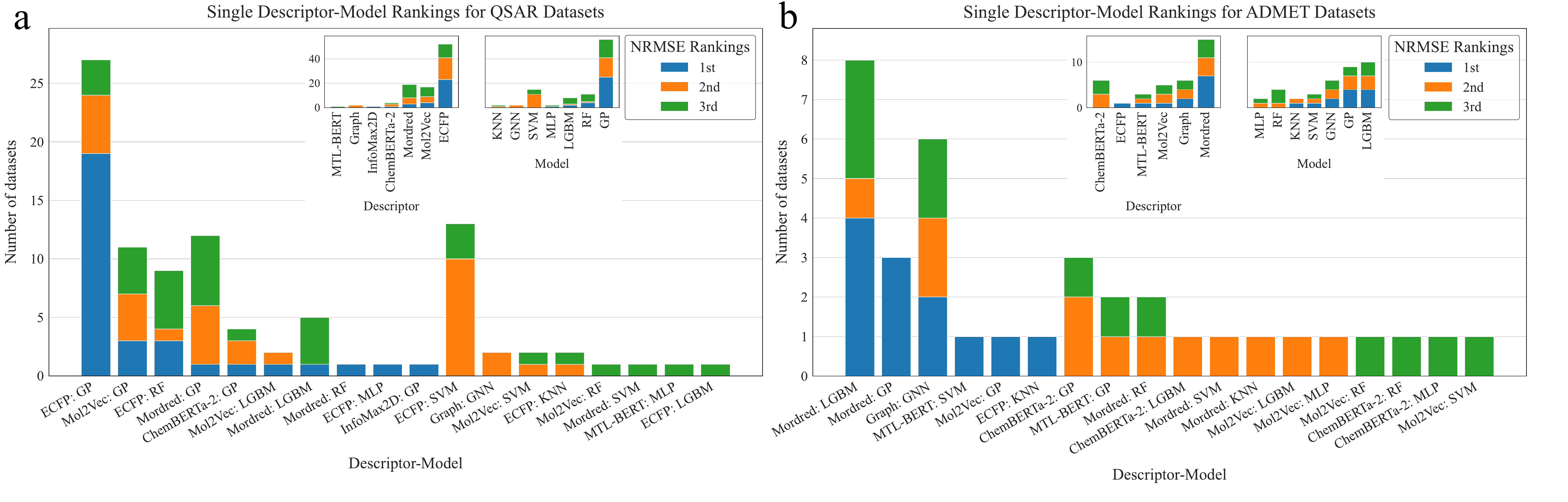}
\caption{Single descriptor-model rankings based on NRMSE for a) QSAR datasets and b) \admet{} datasets. Each count corresponds to the number of datasets on which a descriptor-model combination ranked as 1st, 2nd, or 3rd by NRMSE. The inset illustrates the same but for each descriptor and model independently. Higher numbers indicate better performance.}
\label{fig1:aqsar_badmet}
\end{figure}

The 3 top performing models by NRMSE were identified for each dataset; these rankings are shown in \autoref{fig1:aqsar_badmet}, ordered according to the number of datasets in which they secured the 1st position, followed by the 2nd, and 3rd. Ensemble models are excluded from these results as we observed no statistically significant difference in NRMSE between ensemble and single models (two-sided paired t-tested: $\alpha = 0.05, p = 0.419$, \autoref{fig3:uncert_motivation}b), in agreement with previous literature \cite{ensembles}. 

For QSAR datasets, we found that GPs with ECFP descriptors achieved first place in the majority of datasets (1st in $19/32$ and 2nd or 3rd in $8/32$, \autoref{fig1:aqsar_badmet}a). Irrespective of descriptor, GPs were ranked top in $25/32$ datasets (\autoref{fig1:aqsar_badmet}a inset). The superior performance of GPs for QSAR data in particular is a key finding as many studies that compare classical methods with deep learning techniques neglect to include GPs \cite{molace, couldgnnsbebetter, chithran2020chemberta}. Our analysis also identified SVMs and ECFP as the most frequent 2nd ranked descriptor-model combination, consistent with \citet{molace} (\autoref{fig1:aqsar_badmet}a).

For \admet{} datasets, there was more variability in top performing descriptors and models. Mordred descriptors emerged as the most effective feature space for these datasets, especially when used with LGBMs (\autoref{fig1:aqsar_badmet}b). GNNs and GPs also performed well across these datasets (\autoref{fig1:aqsar_badmet}b). In contrast to QSAR datasets, SVMs and ECFP descriptors appear unable to effectively capture molecular information for these datasets. Additionally, we discovered that NLP-based descriptors (Mol2Vec, MTL-BERT, ChemBERTa-2) performed well on these datasets even when compared to ECFP, in agreement with literature \cite{molace, mtlbert, chithran2020chemberta, couldgnnsbebetter}. Our results suggest that the most effective molecular representation is ECFP for QSAR datasets and dataset dependent for \admet{}.

\subsection{Do specific models demonstrate enhanced performance on particular types of datasets?}
\label{q2_results}

\begin{figure}[t]
\centering
\includegraphics[width=1\textwidth]{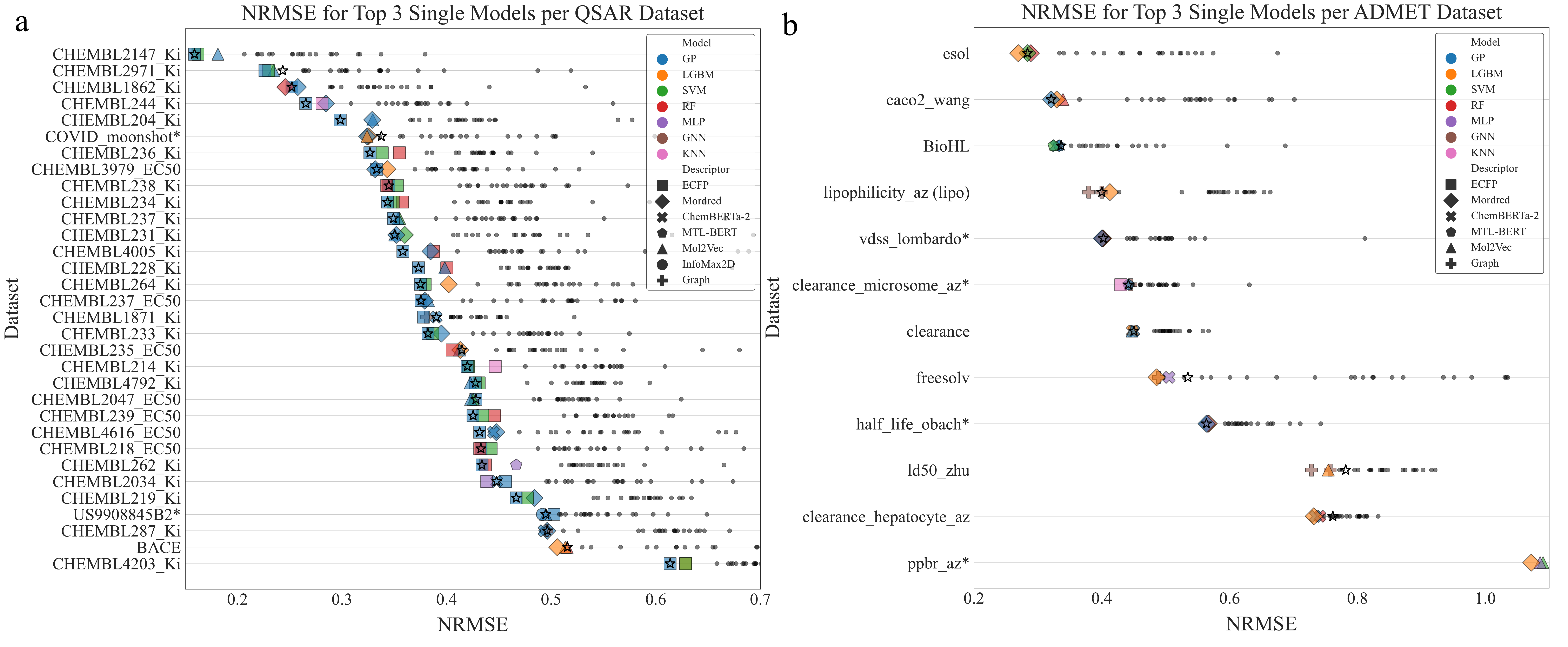}
\caption{NRMSE for the top 3 single models for each dataset, split into a) QSAR datasets, and b) \admet{} datasets. Colour and shape indicate model and descriptor type respectively. Lower NRMSE values (towards the left) indicate better performance. An asterisk (*) indicates that the target values have been log-transformed. The smaller black points correspond to models that did not appear in the top 3, but still fall within the axes limits. Stars ({\mbox{\Large$\smallstar$}}) indicate the NRMSE score for model with the best uncertainty quantification, as measured by NLPD.}
\label{fig2:nrmse}
\end{figure}

\autoref{fig1:aqsar_badmet} provides a measure of model performance, however, this does not describe the magnitude of performance increase in selecting the 1st ranked model, compared to the 2nd or 3rd or indeed models outside of the top 3. \autoref{fig2:nrmse} shows quantitative measurements of model performance, where for each dataset the NRMSE for each trained model is shown with the top 3 models highlighted.

For QSAR datasets, there is a notable performance gap between the top performing models and all others, with only small differences between the best models (\autoref{fig2:nrmse}a). 
For \admet{} datasets (\autoref{fig2:nrmse}b), the best performing model varies by dataset, indicated by a variety of colours amongst the top performing models. NRMSE values are generally greater than for QSAR datasets, suggesting it is more difficult to predict accurately on \admet{} datasets (\autoref{fig2:nrmse}b). As with QSAR datasets, there is generally a notable gap in performance between the best performing models and all others. Where there is a noteworthy difference in performance between the top models, GNNs are often the best performers, e.g.~in the \textit{lipo} (in agreement with \cite{couldgnnsbebetter}) and \textit{ld50\_zhu} datasets, indicating that they are in fact well suited to a specific subset of datasets.
The \textit{ppbr\_az*} dataset appears to be particularly challenging to predict with an NRMSE $ > 1$ (NRMSE $ = 1.07$). However, this is still an improvement over a naïve prediction of the training data mean for all test points, which gives $\text{NRMSE}_\text{mean} = 1.30$.

\subsection{Which models best quantify uncertainty in their predictions?}
\label{q3_results}
As discussed in \cref{methods}, we aim to develop models with good uncertainty estimates, as well as good mean predictions. So in addition to measuring NRMSE, we also measure NLPD, which measures \emph{both} accuracy and uncertainty quantification. Considering this additional metric may complicate drawing conclusions about ``best'' models, as the model with the best mean predictions may have poor uncertainty estimates. To assess whether this is the case, the comparison based on NRMSE in \autoref{fig2:nrmse} has the model with the best NLPD highlighted by a star.
For the QSAR datasets, we see that the model with the best uncertainty quantification always has the best, or one of the best NRMSEs. For some \admet{} datasets (in particular \textit{freesolv}, \textit{ld50\_zhu} and \textit{clearance\_hepatocyte\_az}), the best models by NLPD are outside the top 3.
These demonstrate instances where the NRMSE decreases when selecting the best model by NLPD. Therefore, for certain \admet{} datasets, it might be necessary to strike a balance between the point prediction accuracy and the estimation of uncertainty.

\begin{figure}[t]
\centering
\includegraphics[width=1\textwidth]{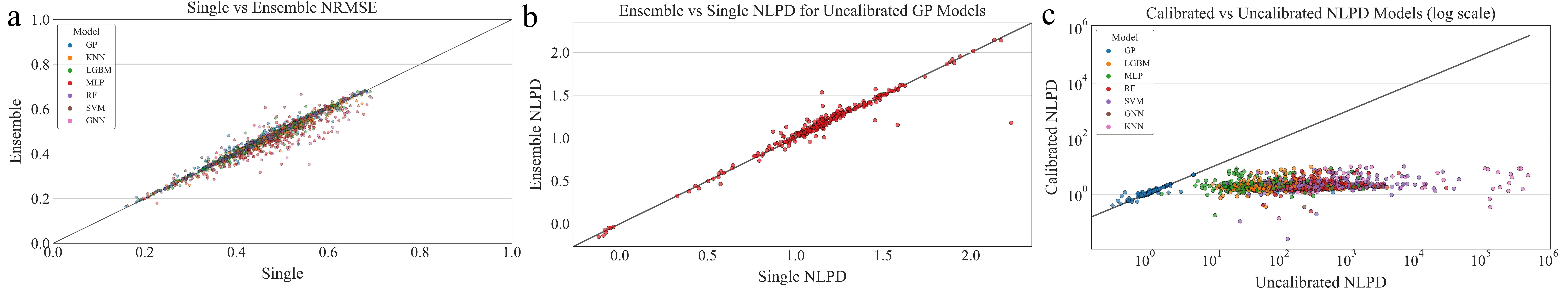}
\caption{a) NRMSE for corresponding single and ensemble models b) NLPD for corresponding single and ensemble GPs. For both a) and b), points below the line indicate that ensembling improved performance compared to a single model c) NLPD for corresponding calibrated and uncalibrated models (single for GPs, ensemble for all others). Points below the line indicate that calibration improved performance compared to leaving a model uncalibrated. Data is shown for all datasets.}
\label{fig3:uncert_motivation}
\end{figure}

Next, we directly compare the NLPD of all models, including ensemble versions of all methods that do not naturally provide uncertainty estimates (i.e.~all methods except GPs). Ensembles are known to improve uncertainty estimates without reducing predictive accuracy (verified in \autoref{fig3:uncert_motivation}a). Ensembled GPs do not improve uncertainty quantification (verified in \autoref{fig3:uncert_motivation}b), leading them to be excluded from the comparisons. 
Since the raw uncertainty values from ensembles are known to contain bias (as shown by \citet{palmer2022calibration}), we also consider calibrated versions of these models, in order to maximally benefit from the signal in ensembles' uncertainty. \autoref{fig3:uncert_motivation}c shows that calibration greatly improves NLPD for ensemble models, while making little difference for GP models.


\begin{figure}[ht]
\centering
\includegraphics[width=1\textwidth]{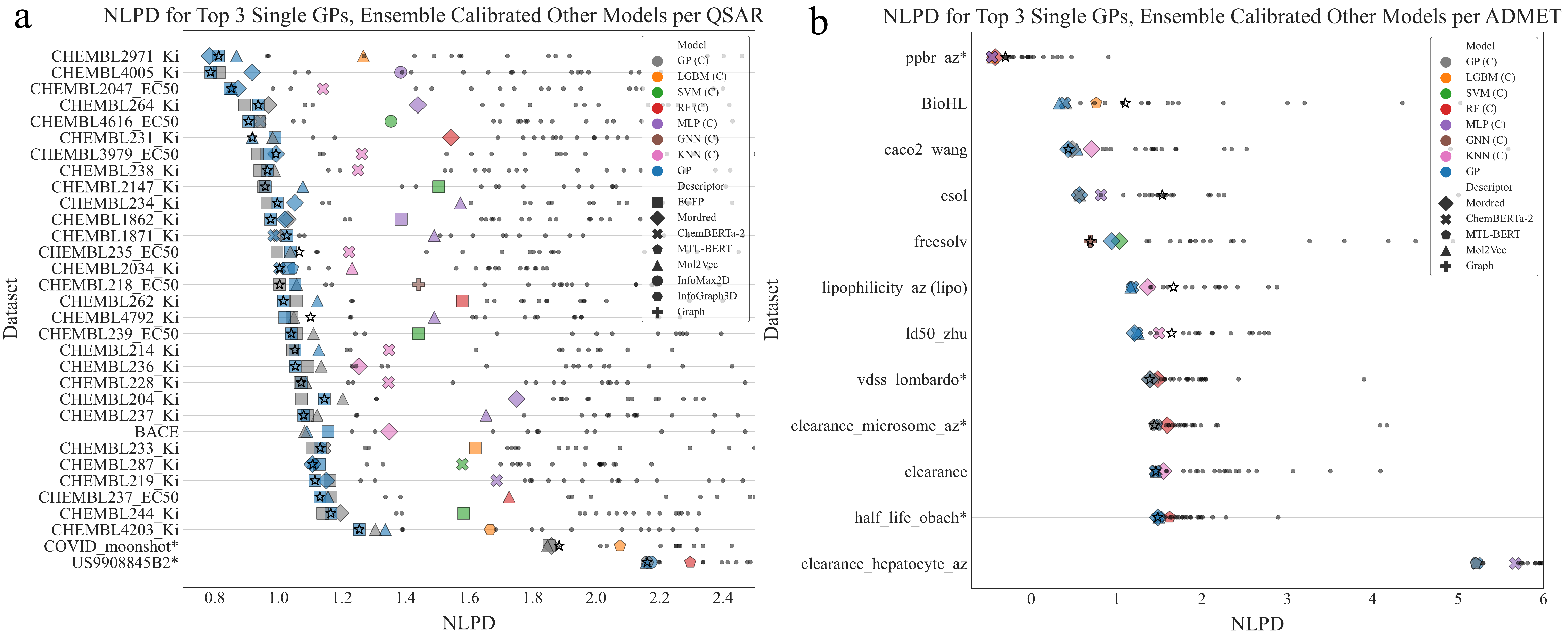}
\caption{NLPD for the top 3 single models for each dataset, split into a) QSAR datasets, and b) \admet{} datasets. Colour indicates model type, and shape descriptor type. Lower NLPD values (towards the left) indicate better performance. An asterisk (*) indicates that the target values have been log-transformed. The smaller black points correspond to models that did not appear in the top 3, but still fall within the axes limits. Stars ({\mbox{\Large$\smallstar$}}) indicate the NLPD for the most accurate model based on NRMSE. The top performing non-GP model by NLPD for each dataset is is also highlighted if all 3 best performing models are GPs. Calibrated models are indicated by (C).}
\label{fig4:nlpd_scores}
\end{figure}

Based on these observations, we compare the NLPD in \autoref{fig4:nlpd_scores} for all ensembled+calibrated models, and calibrated and uncalibrated GPs. For QSAR datasets, both calibrated and uncalibrated GPs provide the most accurate uncertainty estimates, by a wide margin when compared to the best non-GP models (\autoref{fig4:nlpd_scores}a). Selecting a model based on the best NRMSE (highlighted by a star) almost always results in good uncertainty estimation (low NLPD). This is consistent with \autoref{fig2:nrmse}, and is unsurprising given that GPs tend to be the best performers for QSAR datasets in both point and uncertainty prediction (\autoref{fig2:nrmse}a and \autoref{fig4:nlpd_scores}a, respectively). For \admet{} datasets, the results are broadly similar, however there are cases where GPs are outperformed by GNNs and MLPs (\textit{ppbr\_az*} and \textit{freesolv} datasets \autoref{fig4:nlpd_scores}b). Selecting a model based on the best NRMSE does not always result in good uncertainty estimation for these datasets, meaning a compromise may be necessary to balance the quality of point and uncertainty estimates for these property classes. 

\subsection{What are the trade-offs between computational time and performance?}

Drug property prediction datasets are usually small, with the datasets used in this work consisting of between approximately 50 and 3500 data points. Limitations on training and inference time depend on the practitioner's personal requirements. Nevertheless, we provide an overview of featurisation, training and inference time for the descriptors and models used here as a general guide as calculated on our hardware (optimised using GPU acceleration for both training and inference for GNNs, GPs and MLPs, CPU parallelisation for LGBMs and RFs and single threaded CPU otherwise). Computation of the following descriptors in seconds took: $10^0$ (ECFP), $10^1$ (ChemBERTa-2, MTL-BERT, Mol2Vec, InfoMax2D, InfoGraph3D, Graph), $10^3$ (Mordred). Training of the following single models in seconds took: $10^0$ (KNN, MLP), $10^1$ (GP, LGBM, SVM), $10^3$ (RF, GNN). Inference using the following single models in seconds took: $10^{-2}$ (GP, MLP), $10^{-1}$ (LGBM, RF, KNN), $10^0$ (RF, GNN), $10^1$ (SVM). Detailed summaries of runtimes are given in \autoref{fig:compute_time}.

\section{Conclusions and Future Directions}
Our comprehensive evaluation of machine learning methods on a broad selection of benchmark datasets aimed to address the inconsistencies and gaps in the literature for drug property regression tasks. Our results demonstrated that the effectiveness of a particular model is dependent on the property class and the feature descriptors used.

The most striking finding was the superior performance of GPs on QSAR datasets, particularly in conjunction with ECFP descriptors. Despite being often overlooked in benchmarking studies, GPs not only excelled in point prediction but also provided the most accurate uncertainty estimates. This underlines the importance of considering GPs as viable models for QSAR datasets. However, for \admet{} datasets, the choice of the optimal model was less clear. The results suggested that GNNs can lead to very good results and that LGBMs and GPs perform well with Mordred descriptors. For \admet{} datasets, there was a notable trade-off between point prediction accuracy and uncertainty estimation as improving uncertainty decreased accuracy and vice versa.

Some notable limitations of our study were that we did not investigate more complex ensemble methods, different calibration techniques, or pre-training followed by fine-tuning. Moreover, we did not explore other methods for uncertainty estimation, such as hyperplane distance for SVMs or variance in Random Forests. For deep learning models, techniques such as dropout, weight priors, or Stein Variational Gradient Descent could also be investigated.

In conclusion, our study highlights the importance of carefully selecting models based on the specific property at hand and the necessity of uncertainty quantification in drug property prediction tasks. Ideally, this approach requires an automatic framework for model training, testing and selection in order to deliver the best performance for a specific dataset and property class. Future work should aim to expand on these findings, exploring new models, calibration methods, and uncertainty estimation techniques. This would help to further accelerate, de-risk and facilitate more effective decision-making in the drug discovery process.

\bibliography{references}
\bibliographystyle{plainnat}

\newpage
\appendix
\setcounter{figure}{0}
\renewcommand\thefigure{A.\arabic{figure}}
\setcounter{table}{0}
\renewcommand{\thetable}{A.\arabic{table}}

\section{Appendix}

\subsection{Datasets Summary}
\label{app:datasets_summary}

\begin{table}[h]
\centering
\caption{Dataset Summary (44 total). Datasets where the target values have been log-transformed are suffixed with an asterisk (*).}
\label{tab:datasets_summary}
\resizebox{\columnwidth}{!}{%
\begin{tabular}{@{}llllllll@{}}
\toprule
Dataset                         & Property               & Train/Test Split    & Dataset Size & Train Size & Test Size & Source          & Reference \\ \midrule
BACE                            & Affinity                   & DeepChem & 1513         & 1361       & 152       & DeepChem        & \cite{deepchem}         \\
BioHL                           & \admet (Metabolism)   & Opera    & 150          & 112        & 38        & OPERA           & \cite{mansouri2018opera}        \\
caco2\_wang                     & \admet (Absorption)   & TDC (Scaffold)     & 906          & 725        & 181       & TDC             & \cite{Huang2021tdc}        \\
CHEMBL1862\_Ki                  & Affinity               & ACE      & 794          & 633        & 161       & MoleculeACE    & \cite{molace}        \\
CHEMBL1871\_Ki                  & Affinity               & ACE      & 659          & 525        & 134       & MoleculeACE    & \cite{molace}        \\
CHEMBL2034\_Ki                  & Affinity               & ACE      & 750          & 598        & 152       & MoleculeACE    & \cite{molace}        \\
CHEMBL204\_Ki                   & Affinity               & ACE      & 2754         & 2201       & 553       & MoleculeACE    & \cite{molace}        \\
\bottomrule
\end{tabular}%
}
\end{table}

\FloatBarrier
\newpage

\begin{landscape}
\subsection{Numerical Results for Best Performing Models}
\scriptsize
\begin{longtable}[h]{@{}
p{2.2cm} 
p{0.8cm} 
p{0.8cm} 
p{0.6cm} 
p{1.5cm}   
p{1.1cm} 
p{0.8cm} 
p{0.8cm}   
p{0.8cm} 
p{0.8cm}   
p{0.9cm} 
p{1.2cm} 
p{1.6cm}   
p{0.8cm}   
p{1.0cm}   
@{}
}
\caption[]{Full results of top 3 performing models by dataset if ranked by best (lowest) NRMSE or by best (lowest) NLPD. Datasets where the target values have been log-transformed are suffixed with an asterisk (*). Calibrated models are indicated by (C). All results are given to 4 significant figures, except hyperparameter optimisation and training time, which are given to 2 significant figures.} 
\label{tab:experimental_results} \\
\toprule
Dataset & Property Class & Best By & Rank &   Descriptor &     Model &      MSE &    RMSE &   NRMSE & Pearson R &     NLPD & Featurisation (s) & Hyperparameter Optimisation (s) & Training (s) & Inference (s) \\
\midrule
\endfirsthead
\toprule
Dataset & Property Class & Best By & Rank &   Descriptor &     Model &      MSE &    RMSE &   NRMSE & Pearson R &     NLPD & Featurisation (s) & Hyperparameter Optimisation (s) & Training (s) & Inference (s) \\
\midrule
\endhead
\midrule
\endfoot

\bottomrule
\endlastfoot
BioHL & \admet & NRMSE & 1st &     MTL-BERT &       SVM &   0.1118 &  0.3343 &  0.3244 &    0.9034 &    18.71 &                13 &                            0.35 &     0.009481 &      0.006161 \\
             &      &      & 2nd &  ChemBERTa-2 &        GP &   0.1147 &  0.3386 &  0.3286 &    0.8977 &   0.3944 &               1.4 &                              68 &        1.586 &       0.00407 \\
             &      &      & 3rd &     MTL-BERT &        GP &   0.1177 &  0.3431 &  0.3329 &    0.9001 &   0.4507 &                13 &                              74 &       0.5088 &      0.004628 \\
             &      & NLPD & 1st &      Mol2Vec &        GP &     0.12 &  0.3465 &  0.3362 &    0.8967 &   0.3263 &               2.5 &                              61 &       0.4085 &      0.003968 \\
             &      &      & 2nd &  ChemBERTa-2 &        GP &   0.1147 &  0.3386 &  0.3286 &    0.8977 &   0.3944 &               1.4 &                              68 &        1.586 &       0.00407 \\
             &      &      & 3rd &      Mol2Vec &    GP (C) &   0.1263 &  0.3554 &  0.3449 &    0.8852 &    0.408 &               2.5 &                              61 &        1.503 &      0.007095 \\
caco2\_wang & \admet & NRMSE & 1st &      Mordred &        GP &   0.1254 &  0.3542 &  0.3207 &    0.8602 &   0.4306 &               440 &                              93 &        1.087 &       0.01426 \\
             &      &      & 2nd &      Mordred &      LGBM &   0.1322 &  0.3637 &  0.3293 &    0.8465 &    6.149 &               440 &                             130 &        2.945 &       0.01665 \\
\end{longtable}
\end{landscape}

\FloatBarrier
\newpage

\subsection{Runtime Measurements}
\begin{figure}[h]
    \centering
    \begin{subfigure}{\textwidth}
        \centering
        \includegraphics[width=\linewidth]{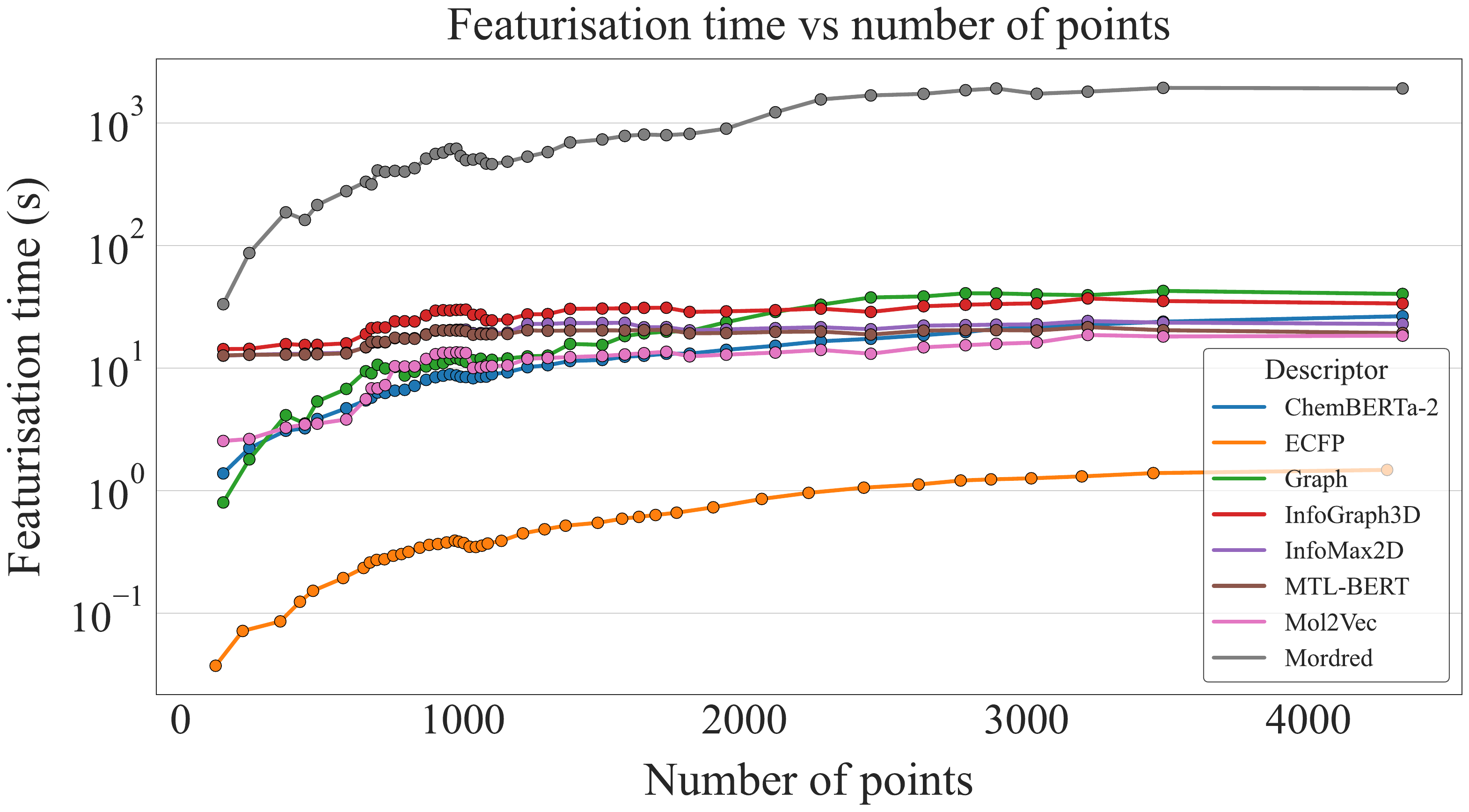}
        \subcaption{}
        \label{fig:feat_time}
    \end{subfigure}
    \vspace*{1cm}
    \\
    \begin{subfigure}{\textwidth}
        \centering
        \includegraphics[width=\linewidth]{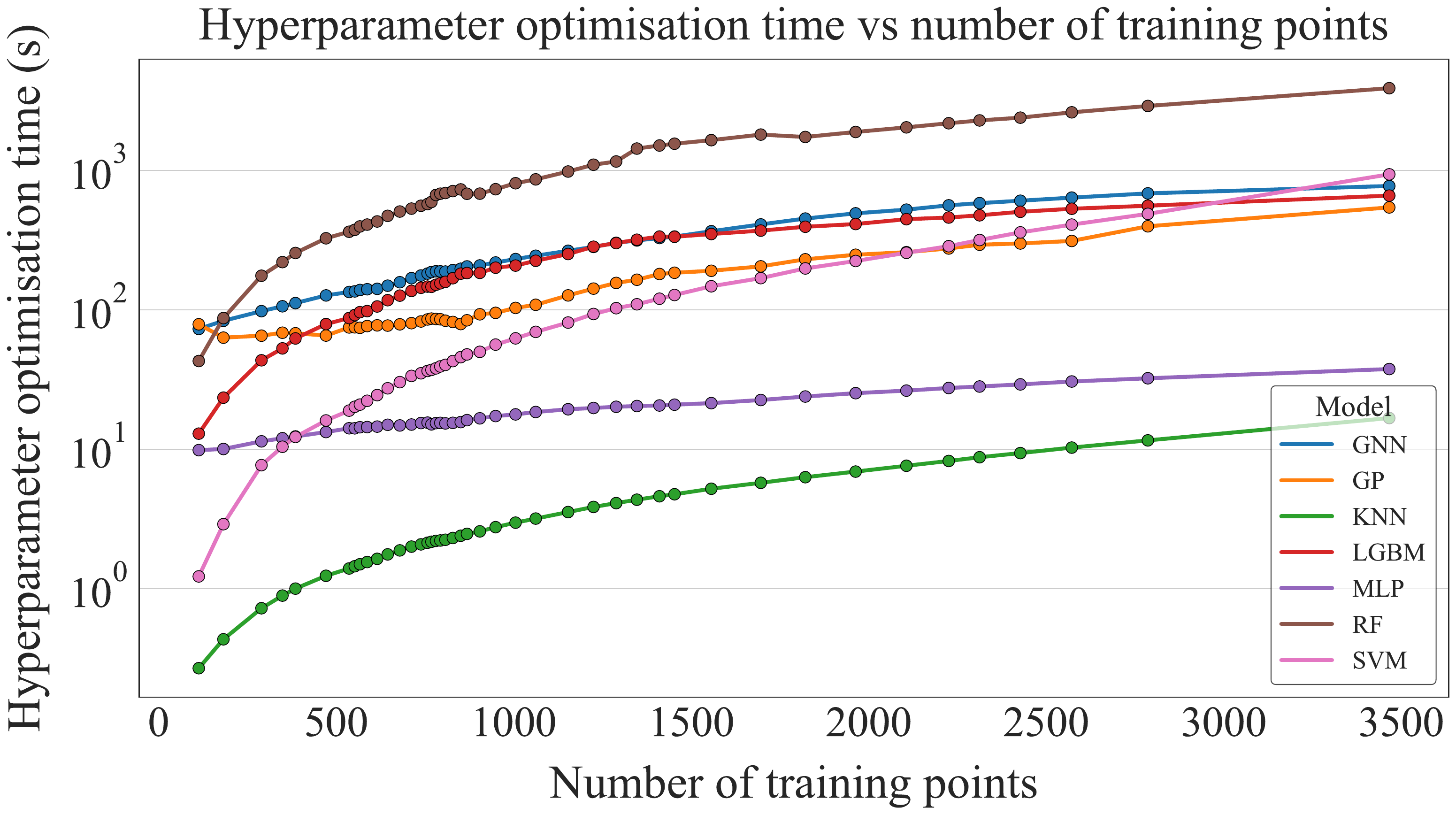}
        \subcaption{}
        \label{fig:hypopt_time}
    \end{subfigure}
\end{figure}

\begin{figure}[h]
    \ContinuedFloat
    \centering
    \begin{subfigure}{\textwidth}
        \centering
        \includegraphics[width=\linewidth]{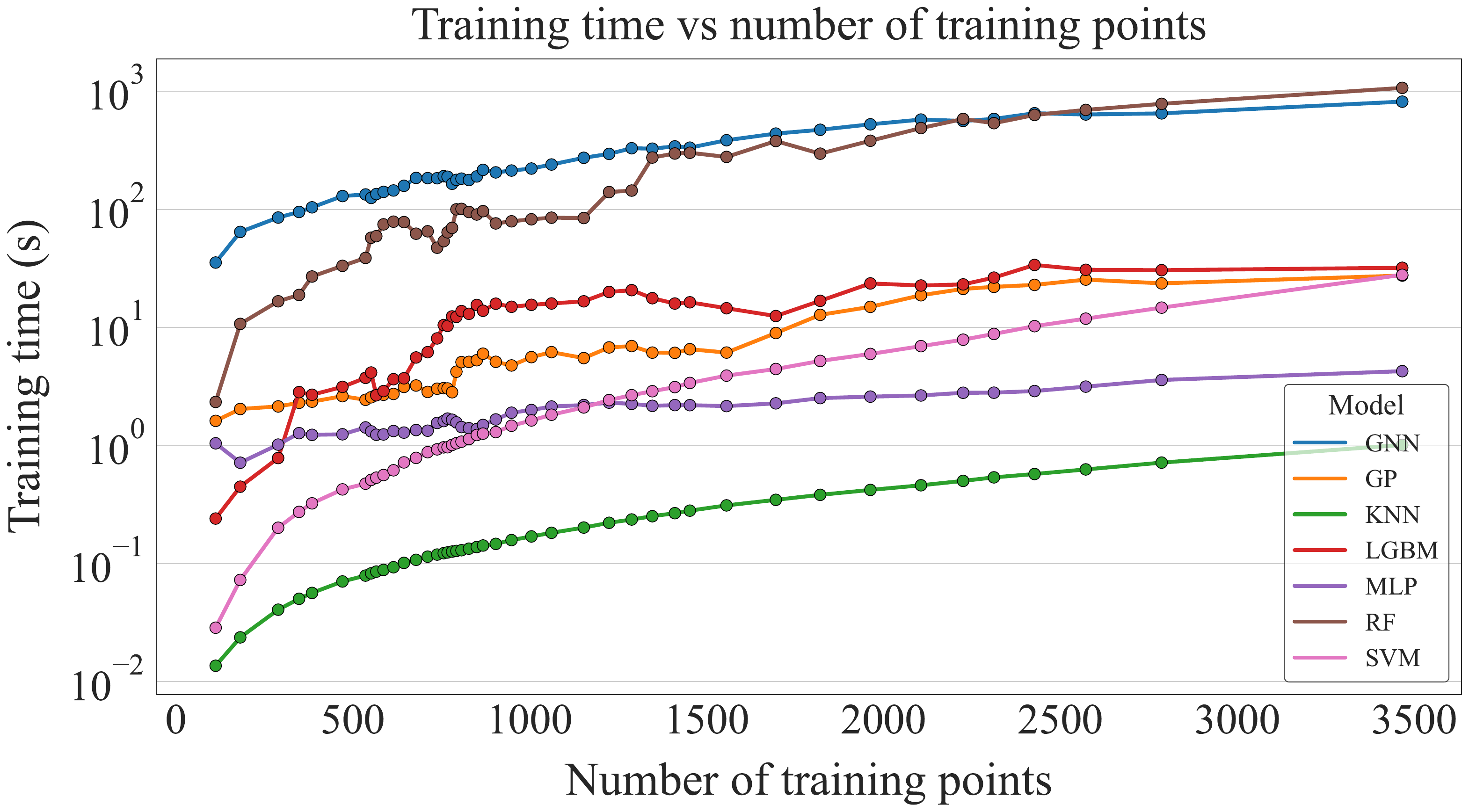}
        \subcaption{}
        \label{fig:train_time}
    \end{subfigure}
    \vspace*{1cm}
    \\
    \begin{subfigure}{\textwidth}
        \centering
        \includegraphics[width=\linewidth]{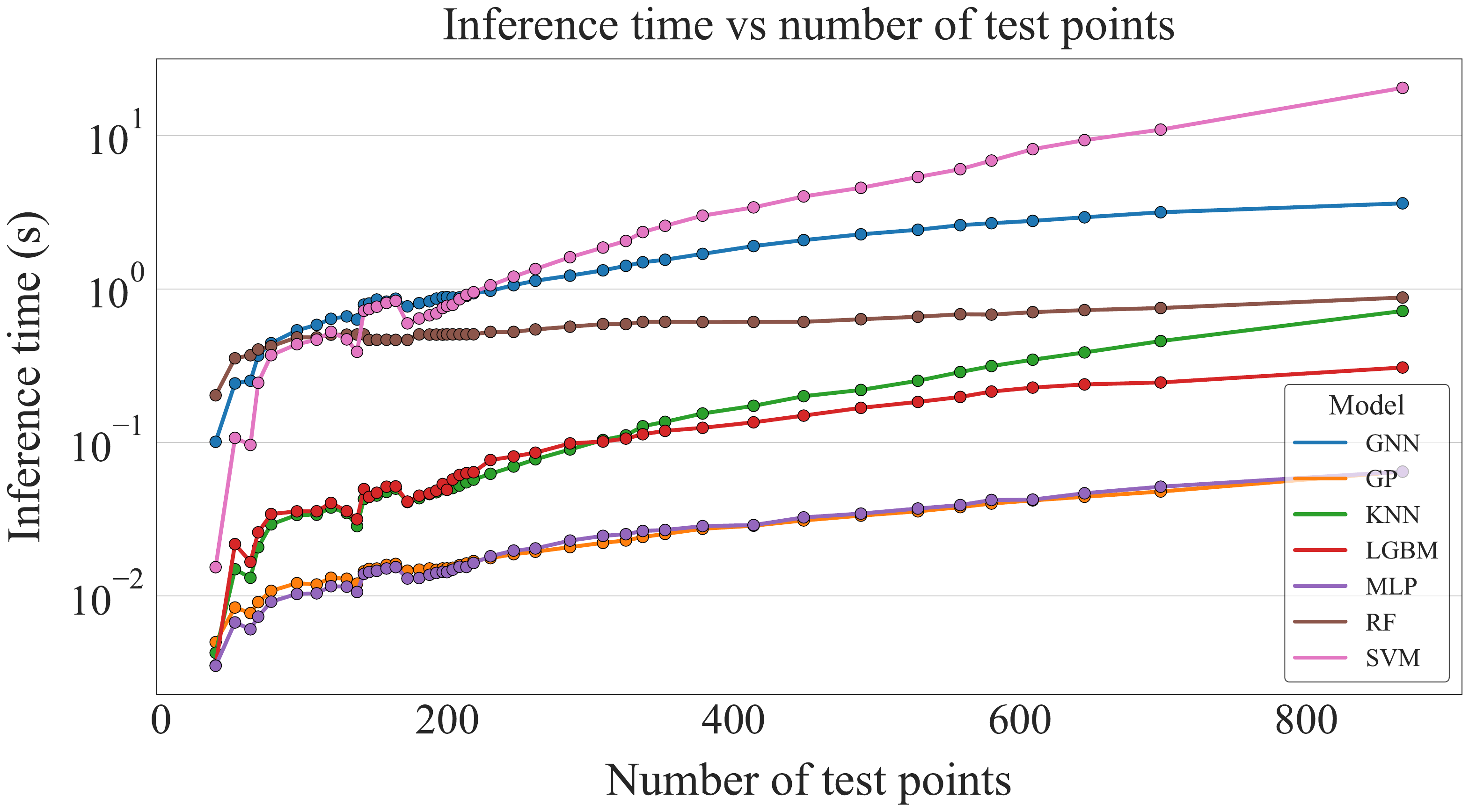}
        \subcaption{}
        \label{fig:inf_time}
    \end{subfigure}
    \caption{Featurisation, hyperparameter optimisation, training and inference runtimes for single uncalibrated models by number of input points, split by descriptor (\subref{fig:feat_time}) or model, (\subref{fig:hypopt_time}), (\subref{fig:train_time}), (\subref{fig:inf_time}). For (\subref{fig:hypopt_time}), (\subref{fig:train_time}) and (\subref{fig:inf_time}), points represent the maximum training time across all descriptors. In all cases, a rolling mean with window size 5 is applied point-wise for smoothing. The following descriptors were calculated on GPU: ChemBERTa-2, MTL-BERT, InfoMax2D, InfoGraph3D and the remainder on CPU. The following models were trained on GPU: Graph, GP, MLP. Remaining models were trained on CPU, with LGBM and RF trained using multiprocessing over 4 cores. The GPU used was an Nvidia T4.}
    \label{fig:compute_time}
\end{figure}

\FloatBarrier
\newpage

\subsection{Hyperparameter Optimisation Search Spaces}
\begin{table}[h]
    \caption{Hyperparameter search spaces  by model. In the case of ensemble models, the same hyperparameters were used for all members. A random search was applied using a search space size of 20 and 3-fold cross validation for each hyperparameter combination. The final model was retrained with the optimal parameters on the whole training dataset.}
    \vspace{0.5cm}
    \scriptsize
    \centering
    \begin{tabular}{lll}
    \toprule
    Model & Hyperparameter & Search Values \\
    \midrule
        SVM & C &  1, 10, 100, 1000 \\ 
              & epsilon &  0.1 \\ 
              & gamma &  1.0e-06, 1.0e-05, 0.0001, 0.001, 0.01, 0.1 \\ 
              & kernel &  RBF \\ 
        \midrule
        RF & n\_estimators &  100, 250, 500, 1000 \\ 
        \midrule
        GP & kernel &  RBF, Matern, RQ, Tanimoto \\ 
              & n\_iter &  10, 100, 1000 \\ 
              & lr &  0.01, 0.1, 1 \\ 
        \midrule
        LGBM & n\_estimators &  100, 200, 500, 1000 \\ 
              & max\_depth &  3, 4, 5, 6, 7 \\ 
              & learning\_rate &  0.01, 0.05, 0.1, 0.2, 0.3 \\ 
              & subsample &  0.4, 0.45, 0.5, 0.55, 0.6, 0.65 \\ 
              & colsample\_bytree &  0.5, 0.6, 0.7, 0.8, 0.9, 1.0 \\ 
              & reg\_alpha &  0, 0.1, 1, 5, 10 \\ 
              & reg\_lambda &  0.1, 1, 10, 100 \\ 
              & min\_child\_weight &  1, 3, 5 \\ 
              & num\_leaves &  10, 20, 40, 80, 200, 500 \\ 
              & min\_child\_samples &  5, 10, 20, 40, 100 \\ 
              & tree\_learner &  feature \\ 
        \midrule
        KNN & n\_neighbors &  3, 5, 11, 21 \\ 
              & weights &  uniform, distance \\ 
        \midrule
        MLP & batch\_size &  128, 256, 512 \\ 
              & learning\_rate &  0.0001, 0.0005, 0.001 \\ 
              & mlp\_dropout &  0.0, 0.1, 0.2, 0.3, 0.4, 0.5 \\ 
              & mlp\_hidden &  128, 256, 512, 1024 \\ 
              & mlp\_n\_layers &  2, 3, 4 \\ 
        \midrule
        A-FP & batch\_size &  128, 256, 512 \\ 
              & learning\_rate &  1e-5, 0.0001, 0.001 \\ 
              & dropout &  0.0, 0.1, 0.2, 0.3, 0.4, 0.5 \\ 
        \midrule
        GAT & batch\_size &  128, 256, 512 \\ 
              & learning\_rate &  1e-5, 0.0001, 0.001 \\ 
              & dropout &  0.0, 0.1, 0.2, 0.3, 0.4, 0.5 \\ 
        \midrule
        GCN & batch\_size &  128, 256, 512 \\ 
              & learning\_rate &  0.0001, 0.0005, 0.001 \\ 
              & graph\_conv\_layers &  16, 16, 32, 32, 64, 64 \\ 
              & batchnorm &  True, False \\ 
              & dropout &  0.0, 0.2, 0.5 \\ 
              & predictor\_dropout &  0.0, 0.2, 0.5 \\ 
        \midrule
        DMPNN & batch\_size &  128, 256, 512 \\ 
              & learning\_rate &  1e-5, 0.0001, 0.001 \\ 
              & enc\_dropout\_p &  0.00, 0.2, 0.5 \\ 
              & ff\_dropout\_p &  0.00, 0.2, 0.5 \\
    \bottomrule 
    \end{tabular}
\end{table}

\FloatBarrier

\end{document}


\maketitle

\begin{abstract}
Predicting drug properties is a vital step in drug development, as it helps to de-risk assets before clinical trials, which is extremely expensive. Interest from the Machine Learning community has led to the release of various benchmark datasets and numerous proposed methods. However, it remains unclear for practitioners which method or approach is most suitable, as different papers benchmark on different datasets and methods, leading to varying conclusions. Our large-scale empirical study links together numerous earlier works by examining the union of datasets and methods. This offers a comprehensive overview of the existing problem types, datasets, and their interactions with different methods. We emphasize the importance of uncertainty quantification and the cost of running methods in the drug development decision-making cycle. Our study demonstrates that the ability to trade off cost and uncertainty allows practitioners to better decide on the most appropriate method. We discover that the best method depends on the dataset, and that engineered features with classical ML methods (SVM, GP) can outperform deep learning. This highlights that practitioners do not yet have a straightforward, black-box procedure to rely on, and sets the precedence for creating practitioner-relevant benchmarks. Deep learning approaches must meet this target to become the practical method of choice in drug property prediction.
\end{abstract}

\section{Introduction}

\section{Motivation}
\label{sec:goals}
Give a broad overview of drug property prediciton, and why it's useful. What are we trying to achieve? Broader context. Acknowledge that good point predictions are good. Good place to signpost that uncertainty is important (because decision-making is involved?) Perhaps discuss cost of methods as well.

\begin{enumerate}
    \item \textbf{RQ1}: Which machine learning model, across the broad spectrum of existing models, consistently exhibits superior performance in a variety of tasks?
    \item \textbf{RQ2}: Are there specific models that demonstrate enhanced performance when matched with particular types of datasets?
    \item \textbf{RQ3}: To what extent are the considered models capable of effectively capturing uncertainty?
    \item \textbf{RQ4}: What are the trade-offs among computational time, financial cost, and performance efficacy in the application of these models?    
\end{enumerate}

\section{Drug Property Prediction Problems and Datasets}

Numerous public benchmarks are available for developing models in property prediction for ADME(T) and QSAR, with the most prominent ones stemming from the Therapeutics Data Commons (TDC) ~\cite{Huang2021tdc, Huang2022artificial} and MoleculeNet~\cite{wu2018moleculenet}. In our study, we specifically focus on regression tasks from these benchmarks, as they hold several advantages over classification tasks in real-world scenarios:

\begin{enumerate}
  \item Many molecular property prediction classification tasks are derived from regression problems, leading to information loss during thresholding.
  \item Classification outputs do not allow for prioritization, which is a vital component of the drug discovery process.
  \item When converting a problem into a classification task, practitioners must choose a threshold, which is subjective and may differ among individuals within the same organization.
\end{enumerate}

Consequently, our exclusive focus on regression tasks within these public benchmarks aligns with their heightened importance in real-world applications. Molecular property prediction often grapples with the issue of activity cliffs, where minor alterations in molecular structure can cause significant property changes by several orders of magnitude. To address this, we have included several activity cliff datasets, curated by van Tilborg \textit{et al.}~\cite{molace} from ChemBLE, in our benchmark.

For all benchmarks, we adhered to the default split as reported in the original benchmark. We conducted train/test splits using four distinct methods: random, scaffold, minmax, and activity cliff. A detailed overview of the train and test sizes is available in Table Xa of the appendix.

\section{Current Methods}
Overview of current methods, and claims made in the papers.

\section{Challenges in Scalable Real-World Application}
Raise questions about claims in previous section (perhaps merge with previous section). Discuss how our 3 goals from \cref{sec:goals} are not yet met (knowing what method performs best pointwise, UQ, cost). Raise broad questions we aim to discuss.

\section{Experiments}
\begin{itemize}
    \item List specific, answerable questions that we wish to answer.
    \item Each experiment should answer a question.
\end{itemize}

\section{Conclusions}

\begin{ack}
Use unnumbered first level headings for the acknowledgments. All acknowledgments
go at the end of the paper before the list of references.
\end{ack}

\bibliography{references}